\documentclass[prl,superscriptaddress,twocolumn,showpacs]{revtex4}

\usepackage{mathptmx}
\usepackage{helvet}
\usepackage{graphicx,fancyhdr}
\usepackage{dashbox,color}%
\usepackage{bm}
\usepackage{sidecap}%
\definecolor{lightblue}{rgb}{0.6,0.9,1}
\definecolor{myrefblue}{rgb}{0.1,0.6,1}
\definecolor{myblue}{rgb}{0,0,0}
\definecolor{nmat}{rgb}{0.7,0.04,0.26}

\newcommand{\sfaof}{\mbox{SmFeAsO$_\mathrm{1-x}$F$_\mathrm{x}$}}
\newcommand{\pfaof}{\mbox{PrFeAsO$_\mathrm{1-x}$F$_\mathrm{x}$}}
\newcommand{\pfaofdop}{\mbox{PrFeAsO$_\mathrm{0.6}$F$_\mathrm{0.35}$}}

\newcommand{\bkfaopt}{\mbox{Ba$_{0.68}$K$_{0.32}$Fe$_2$As$_2$}}

\begin{document}

\title{Intrinsic charge dynamics in high-$T_\mathrm{c}$ $A$FeAs(O,F) superconductors}

\author{A.~Charnukha}
\email[E-mail:]{acharnukha@ucsd.edu}
\affiliation{Physics Department, University of California--San~Diego, La Jolla, CA 92093, USA}
\affiliation{Leibniz Institute for Solid State and Materials Research, IFW, 01069 Dresden, Germany}
\affiliation{Max Planck Institute for Solid State Research, 70569 Stuttgart, Germany}

\author{D.~Pr\"opper}
\affiliation{Max Planck Institute for Solid State Research, 70569 Stuttgart, Germany}

\author{N.~D.~Zhigadlo}
\affiliation{Laboratory for Solid State Physics, ETH Zurich, CH-8093 Zurich, Switzerland}
\affiliation{Department of Chemistry and Biochemistry, University of Bern, Freiestrasse 3, CH-3012 Bern, Switzerland}

\author{M. Naito}
\affiliation{Department of Applied Physics, Tokyo University of Agriculture and Technology, Koganei, Tokyo 184-8588, Japan}
\affiliation{TRIP, Japan Science and Technology Agency (JST), Chiyoda, Tokyo 102-0075, Japan}

\author{M.~Schmidt}
\author{Z.~Wang}
\author{J.~Deisenhofer}
\author{A.~Loidl}
\affiliation{Experimental Physics V, Center for Electronic Correlations and Magnetism, Institute of Physics,
University of Augsburg, D-86159 Augsburg, Germany}

\author{B.~Keimer}
\author{A.~V.~Boris}
\affiliation{Max Planck Institute for Solid State Research, 70569 Stuttgart, Germany}

\author{D.~N.~Basov}
\affiliation{Physics Department, University of California--San~Diego, La Jolla, CA 92093, USA}
\affiliation{Department of Physics, Columbia University, New York, New York 10027, USA}

\begin{abstract}

We report the first determination of the in-plane complex optical conductivity of 1111~high-$T_\mathrm{c}$ superconducting iron oxypnictide single crystals \pfaofdop\ and thin films \sfaof\ by means of {\it bulk-sensitive} conventional and micro-focused infrared spectroscopy, ellipsometry, and time-domain THz transmission spectroscopy. A strong itinerant contribution is found to exhibit a dramatic difference in coherence between the crystal and the film. Using extensive temperature-dependent measurements of THz transmission we identify a previously undetected 2.5-meV collective mode in the optical conductivity of SmFeAs(O,F), which is strongly suppressed at $T_\mathrm{c}$ and experiences an anomalous $T$-linear softening and narrowing below $T^*\approx110\ \textrm{K}\gg T_\mathrm{c}$. The suppression of the infrared absorption in the superconducting state reveals a large optical superconducting gap with a similar gap ratio $2\Delta/k_\mathrm{B}T_\mathrm{c}\approx7$ in both materials, indicating strong pairing.
\end{abstract}


\maketitle

Almost a decade of intensive research into the phenomenology of high-transition-temperature (high-$T_\mathrm{c}$) iron-based superconductors~\cite{kamihara} has revealed that the $T_\mathrm{c}$ in the vast majority of these compounds is limited to below about $40\ \textrm{K}$ (Ref.~\onlinecite{PhysRevB.83.214520}). Two notable exceptions to this rule are the oxypnictides of the 1111-type $A$FeAs(O,F) family ($A$ stands for a rare-earth metal) with $T_\mathrm{c}$'s up to about $55\ \textrm{K}$ (Ref.~\onlinecite{Ren_discovery_Sm1111}) and the monolayer FeSe grown on SrTiO$_3$~\cite{MonolayerFeSe_65K_2013,STM_monolayerFeSe-STO_2015,FeSe-STO-MgO_etching_2016,0953-2048-29-12-123001} with a $T_\mathrm{c}\approx65\ \textrm{K}$. It is now clear, that in the latter case the abnormally high $T_\mathrm{c}$ is afforded not only by the electronic structure and interactions inherent in the iron pnictides and chalcogenides~\cite{Johnston_Review_2010,0034-4885-74-12-124512,0953-8984-26-25-253203,RevModPhys.87.855,NatPhys_Dai_Dagotto_ReviewMagnPnictides} but also by additional, extrinsic, interfacial interactions of itinerant carriers in FeSe with the SrTiO$_3$ substrate~\cite{Shen_MonolayerFeSe_2014,PhysRevLett.118.067002,1367-2630-18-2-022001,0953-2048-29-5-054009}. In the absence of the latter, the maximum $T_\mathrm{c}$ attainable in monolayer FeSe was found to only reach the aforementioned limit of $\sim40\ \textrm{K}$~\cite{Monolayer_FeSe_dopingdepARPES2015,FeSe-STO-MgO_etching_2016,PhysRevLett.118.067002}.

\begin{figure}[!b]
\includegraphics[width=\columnwidth]{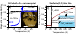}
\caption{\label{fig:char}~(a) Temperature dependence of the ac magnetic susceptibility of \pfaofdop\ cooled in a zero (blue line) and a 10~Oe (black line) magnetic field, subsequently measured in a 10~Oe field in both cases. The deviation from perfect diamagnetism $\chi=-1$ is due to the geometric factor. (inset) Mass and dimensions of the sample used in the present optical study.~(b) Temperature dependence of the dc~resistivity of the optimally doped SmFeAs(O,F) thin film. (inset) Schematic illustration of the sample geometry and the doping mechanism by fluorine diffusion upon annealing.}
\end{figure}
\begin{figure*}[!t]
\includegraphics[width=\textwidth]{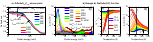}
\caption{\label{fig:opt}~(a) Temperature dependence of the infrared reflectance of \pfaofdop. Vertical dashed line indicates the energy $E_0$ at which reflectance reaches unity.~(b) Temperature dependence of the optical conductivity of SmFeAs(O,F) (solid lines) compared to that of \pfaofdop\ (dotted line).~(c) Temperature dependence of the real part of the dielectric function at various photon energies as indicated in the panel. Vertical dashed line indicates the $T_\mathrm{c}$ determined from the dc resistivity in Fig~\ref{fig:char}b.~(d) Temperature dependence of the real part of the optical conductivity at the same photon energies as in panel~(c). Additional thermal anomaly is visible at $T=T^*$.}
\end{figure*}

These experimental observations emphasize the uniqueness of the superconducting state in the $A\textrm{FeAs(O,F)}$ materials as they reach the 55-K mark unassisted by extrinsic interactions and hold the key to our understanding of the mechanism of high-$T_\mathrm{c}$ superconductivity and further enhancing the superconducting transition temperatures in iron-based compounds. Unfortunately, high-quality single crystals of these materials can only be obtained by a laborious high-pressure growth technique~\cite{PhysRevB.86.214509}, which produces microscopic samples. Small linear dimensions and mass effectively bar a large number of bulk-sensitive experimental techniques from contributing to our knowledge base of iron-oxypnictide phenomenology. After several pioneering optical works on polycrystalline 1111-type samples at the dawn of the iron-pnictide research~\cite{Dubroka:2008id,0295-5075-83-2-27006,boris:027001}, few further attempts have been made at determining the intrinsic optical itinerant response of iron oxypnictides within the superconducting FeAs-layers and its modification in the superconducting state~\cite{PhysRevB.81.224504,PhysRevB.87.180509}. Another major complication is the polar character of the cleaved crystal planes, which leads to excess charge on the sample surface and makes the interpretation of angle-resolved photoemission spectroscopy as well as scanning-tunneling spectroscopy measurements far from trivial~\cite{PhysRevB.82.075135,Yang2011460,Charnukha_NdFeAsOF_FSsingularities2015,Charnukha_SFCAO_2015}. Currently no consensus exists regarding the bulk electronic structure of iron oxypnictides. It is, therefore, imperative to investigate the charge dynamics of these materials and its modification in the superconducting state using a {\it bulk-sensitive} probe of the electronic structure and interactions.

In this Letter, we report the results of a bulk-sensitive broadband optical-spectroscopy study that overcomes the aforementioned materials-related challenges. We use two complementary approaches to shed first direct light onto the bulk charge-carrier response of iron oxypnictides and its modification in the superconducting state. The first approach employs conventional and micro-focused Fourier-transform infrared reflectance spectroscopy as well as micro-focused CCD-based spectroscopic ellipsometry to investigate the intrinsic electrodynamics of microscopic high-pressure--grown~\cite{PhysRevB.86.214509} single crystals of \pfaofdop\ (see Fig.~\ref{fig:char}a) in a wide spectral range from $15\ \textrm{meV}$ to $6\ \textrm{eV}$. The second approach makes use of a unique fluorine-diffusion doping process by means of {\it in situ} annealing after growth of non-superconducting parent SmFeAsO single-crystalline thin films synthesized by state-of-the-art molecular beam epitaxy on a CaF$_2$ substrate and capped by a SmF$_3$ layer~\cite{0953-2048-25-3-035007} (see Fig.~\ref{fig:char}b). This approach has been shown to result in high-quality optimally doped iron-oxypnictide \hbox{SmFeAs(O,F)} thin films with a maximum $T_\mathrm{c}=55\ \textrm{K}$. We have carried out extensive synchrotron-- and thermal-source--based variable-angle-of-incidence spectroscopic ellipsometry as well as time-domain THz transmission spectroscopy measurements on these films in the range from $1\ \textrm{meV}$ to $6.5\ \textrm{eV}$ and at temperatures from $4$~to~$300\ \textrm{K}$. These comprehensive measurements allowed us to extract the complex optical conductivity of the SmFeAs(O,F) thin film from that of the total response of the multilayer structure and access the itinerant-carrier dynamics in this material down to energies well below those afforded by the linear dimension of single-crystalline microcrystals.

The central observations of our work are summarized in Fig.~\ref{fig:opt}. We find high values of the infrared reflectance in the \pfaofdop\ microcrystal (Fig.~\ref{fig:opt}a), indicative of a strong itinerant response. By means of a Drude-Lorentz fit~\cite{PhysRevB.84.174511} we extract the total plasma frequency of $1.4\ \textrm{eV}$ --- on par with many other iron-based superconductors~\cite{0953-8984-26-25-253203} despite the extremely singular band structure of the 1111-type materials~\cite{Charnukha_SFCAO_2015,Charnukha_NdFeAsOF_FSsingularities2015}. This itinerant response reaches a very high degree of coherence at lowest temperatures (quasiparticle scattering rate $\gamma$ of about $5\ \textrm{meV}$) --- a property of materials with low levels of crystalline disorder. Below $T_\mathrm{c}=24\ \textrm{K}$ the infrared reflectance approaches unity below the characteristic energy $E_\mathrm{0}=28\ \textrm{meV}$ indicative of the opening of a nodeless superconducting gap~\cite{Tinkham_superconductivity_1995,RevModPhys.77.721}. Such a high gapping energy is remarkable for a superconductor with $k_\mathrm{B}T_\mathrm{c}\approx2\ \textrm{meV}$.

In the \sfaof\ thin film, analogous Drude-Lorentz decomposition of the optical conductivity (shaded areas in Fig.~\ref{fig:opt}b) reveals an equally strong itinerant response but significantly less coherent (as can also be seen from the direct comparison with the optical conductivity of the \pfaofdop\ microscrystal shown as a dotted line in Fig.~\ref{fig:opt}b). The quasiparticle scattering rate is found to be $150\ \textrm{meV}$ at $300\ \textrm{K}$ and remains unchanged down to lowest temperatures, thus exceeding its \pfaofdop\ counterpart by almost two orders of magnitude.

By virtue of the large surface area of our \sfaof\ thin film we were able to investigate its optical response deep in the infrared regime extending to sub-THz frequencies. This unprecedented for the 1111-type iron oxypnictides spectroscopic access uncovered the existence of a low-energy collective mode, manifested as a broad peak in the optical conductivity centered at $2.5\ \textrm{meV}$ at room temperature. The peak narrows and grows dramatically with decreasing temperature, approaching $15\ \textrm{m}\Omega^{-1}\textrm{cm}^{-1}$ at its maximum --- an order of magnitude larger than the normal-state dc resistivity values of up to $2\ \textrm{m}\Omega^{-1}\textrm{cm}^{-1}$. It then rapidly decreases at lower photon energies to values consistent with the dc transport. The detection of this collective mode requires robust optical access to energies below $3\ \textrm{meV}$ and has not been observed previously in any iron pnictide or chalcogenide.

In the superconducting state, we find a strong signature of a coherent superconducting condensate, manifested in the drastic suppression of the real part of the dielectric function in the THz spectral range (Fig.~\ref{fig:opt}c). The real part of the optical conductivity in Fig.~\ref{fig:opt}d is likewise sensitive to the onset of superconductivity and allows us to extract the superconducting energy gap in what follows. Finally, we discover a distinct temperature scale of $T^*=110\ \textrm{K}\gg T_\mathrm{c}$ (black arrow in Fig.~\ref{fig:opt}d), at which the real part of the optical conductivity displays an additional thermal anomaly.

\begin{figure}[!t]
\includegraphics[width=\columnwidth]{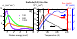}
\caption{\label{fig:mode}~(a) THz conductivity at three different temperatures revealing the thermal evolution of the collective mode. Open circles --- experimental data, solid lines --- fit using two asymmetric Lorentzians on a linear background.~(b) Temperature dependence of the collective mode oscillator strength $f_0$ (black symbols and line), energy $\hbar\omega_0$ (red symbols and line), and width $\hbar\Gamma$ (blue symbols and line) extracted from the fit in panel~(a). Dashed blue line indicates the linear temperature dependence of $\hbar\Gamma$ between $T_\mathrm{c}$ (black dashed line) and $T^*$ (red arrow).}
\end{figure}
\begin{figure}[!b]
\includegraphics[width=\columnwidth]{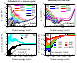}
\caption{\label{fig:sc}~(a,c) Temperature dependence of the infrared optical conductivity of \pfaofdop\ and optimally doped SmFeAs(O,F) thin film, respectively. Hatched area indicates the missing spectral weight in the superconducting state that is transferred into the coherent response of the Cooper-pair condensate at zero energy.~(b) Photon-energy dependence of the far-infrared conductivity of \pfaofdop\ above ($32\ \textrm{K}$, cyan circles) and below ($8\ \textrm{K}$, black circles) $T_\mathrm{c}=24\ \textrm{K}$ normalized to that in the normal state at $24\ \textrm{K}$. Black arrow indicates the photon energy $E_0$ at which optical absorption is completely suppressed (equivalently, reflectance reaches unity) in the superconducting state.~(d) Photon-energy dependence of the far-infrared conductivity of the SmFeAs(O,F) thin film at several temperatures in the superconducting state normalized to that in the normal state at $60\ \textrm{K}$. Vertical dashed line indicates the largest energy of the maximum suppression of the infrared conductivity (consistent with the optical superconducting gap $2\Delta$ in an impure superconductor). Grey line is a fit to the $10\ \textrm{K}$ data using the Mattis-Bardeen expression for the normalized optical conductivity of an impure superconductor in the superconducting state~\cite{PhysRev.111.412}.}
\end{figure}

The low-energy collective mode shows dramatic sensitivity to both $T_\mathrm{c}$ and $T^*$. To demonstrate it, we fit the energy dependence of the real part of the THz conductivity $\sigma_1(\hbar\omega)$ using two asymmetric Lorentzians on a linear background. The results of such a fit are shown in Fig.~\ref{fig:mode}a for three representative temperatures. The excellent quality of the fit allows us to extract the Loretzian parameters for all investigated temperatures with low uncertainty. The temperature dependence of the oscillator strength $f_0$, center energy of the mode $\hbar\omega_0$, and the mode width $\hbar\Gamma$ is plotted in Fig.~\ref{fig:mode}b and clearly reveals the two characteristic temperatures present in this compound: $T_\mathrm{c}$ and $T^*$. The oscillator strength $f_0$ shows a dramatic suppression upon entering the superconducting state below $55\ \textrm{K}$ but reveals no strong anomalies near $T^*$. The mode energy $\hbar\omega_0$ shows the opposite behavior, dropping at $T^*$ with no discernible thermal anomaly at the superconducting transition temperature. The mode width $\hbar\Gamma$ is sensitive to both $T_\mathrm{c}$ and $T^*$.

We hypothesize that the observed collective mode could originate in the quantum critical fluctuations of incommensurate density-wave order. Density-wave fluctuations/order have been found in both the spin~\cite{Drew_Bernhard_coexistence_Sm1111_2009,Drew:2008hp} and, possibly, charge~\cite{Martinelli:2017gf} channel in proximity to the optimally doped iron-oxypnictide superconductors. The hydrodynamic description of these fluctuations indicates that they should manifest themselves as a low-energy collective mode in the optical conductivity of strongly correlated bad metals~\cite{Delacretaz:2016tg}, such as iron-based superconductors~\cite{0953-8984-26-25-253203}. Both the mode energy $\hbar\omega_0$ and width $\hbar\Gamma$ are predicted to exhibit a conspicuous linear temperature dependence, $\hbar\omega_0\sim\hbar\Gamma\sim k_\mathrm{B}T$, analogous to the $T$-linear dc resistivity observed in many unconventional superconductors~\cite{Bruin:2013hc,Hayes:2016bs}. Fig.~\ref{fig:mode}b shows that both $\hbar\omega_0$ and $\hbar\Gamma$ of the collective mode in the SmFeAs(O,F) thin film display a clear linear temperature dependence below $T^*$, consistent with the aforementioned interpretation.

In both the \pfaofdop\ microcrystal and the SmFeAs(O,F) thin film the onset of superconducting coherence is manifested in the transfer of a portion of the infrared spectral weight (hatched areas in Figs.~\ref{fig:sc}a,c) to the dissipationless response at zero frequency according to the Ferrell-Glover-Tinkham sum rule~\cite{PhysRev.109.1398}. This spectral weight corresponds directly to the London penetration depth of a superconductor and in our analysis amounts to $\lambda^\mathrm{Pr}_\mathrm{L}=190\pm100\ \textrm{nm}$ in the microcrystal and a significantly larger $\lambda^\mathrm{Sm}_\mathrm{L}=550\pm50\ \textrm{nm}$ in the thin film.

The signatures of the superconducting optical gap are best revealed in the ratio of the optical conductivity below $T_\mathrm{c}$ to that in the normal state just above $T_\mathrm{c}$: $\tilde{\sigma_1}(\omega)=\sigma^\mathrm{SC}_1(\omega)/\sigma^\mathrm{NS}_1(\omega)$. We examine these ratios for the case of the \pfaofdop\ microcrystal and SmFeAs(O,F) thin film in Figs.~\ref{fig:sc}b,d, respectively. Corresponding to the near-unity reflectance below $E_\mathrm{0}$ in the superconducting state of \pfaofdop\ in Fig.~\ref{fig:opt}a, $\tilde{\sigma_1}(\omega)$ for this material vanishes below the same energy. In a conventional superconductor with a high impurity concentration, the onset of absorption in the superconducting state occurs when the photon energy is sufficient to dissociate a Cooper pair with the binding energy of $2\Delta$~\cite{Tinkham_superconductivity_1995_articlestyle,PhysRev.111.412}. However, we have demonstrated earlier (see Fig.~\ref{fig:opt}a and the corresponding discussion in the text), that \pfaofdop\ microcrystals exhibit a high degree of coherence at low temperatures. In such a clean superconductor, the direct dissociation of a Cooper pair by an incident photon in a two-body process is forbidden by the conservation of energy and momentum. For optical absorption to occur at low temperatures, a quantum of the field mediating the pairing interaction must be excited in addition. If the excitation spectrum of the mediating boson is gapped up to the energy $E_\mathrm{g}$ then absorption becomes allowed above $2\Delta+E_\mathrm{g}$ (Ref.~\onlinecite{PhysRevB.43.473}). In iron-based superconductors the mediating interaction is believed to be of spin-fluctuation origin and indeed has a gapped excitation spectrum in the superconducting state~\cite{PhysRevB.83.214520}, with the spin-gap energy $E_\mathrm{g}$ reaching $2\Delta$~\cite{PhysRevB.88.064504,Charnukha_NaFeCoAs_OpticsARPES}. The combination of multiple Andreev reflection spectroscopy~\cite{Andreev_two_gaps_FeSe_2013} and powder inelastic neutron scattering~\cite{PhysRevB.82.172508} clearly demonstrate that the gap energy in the family of 1111-type materials is $E_\mathrm{g}\approx2\Delta$. Therefore, optical absorption in the superconducting state is expected to occur at an energy of $2\Delta+E_\mathrm{g}\approx4\Delta$, which in the present case results in $\Delta\approx7\ \textrm{meV}$ and a gap ratio of $2\Delta/k_\mathrm{B}T_\mathrm{c}\approx7$, in a good agreement with the largest values found in the pnictides in general~\cite{PhysRevB.83.214520} and, more importantly, in the materials of the same family via ARPES~\cite{Charnukha_NdFeAsOF_FSsingularities2015}. Signatures of mediating-boson--assisted absorption in the infrared conductivity have been previously identified in $\textrm{Ba}_{0.68}\textrm{K}_{0.32}\textrm{Fe}_2\textrm{As}_2$, $\textrm{BaFe}_2(\textrm{As}_{0.67}\textrm{P}_{0.33})_2$, and $\textrm{NaFe}_{0.978}\textrm{Co}_{0.022}\textrm{As}$ in Refs.~\onlinecite{PhysRevB.84.174511,PhysRevLett.109.027006,Charnukha_NaFeCoAs_OpticsARPES}, respectively.

Similarly to the case of \pfaofdop, $\tilde{\sigma_1}(\omega)$ below $T_\mathrm{c}$ in the optimally doped SmFeAs(O,F) thin film reveals a plateau below an energy of about $33\ \textrm{meV}$ (see Fig.~\ref{fig:sc}c), albeit the absorption does not vanish completely at any photon energy and a sizable residual optical conductivity is present (this observation is consistent with the previous steady-state and ultrafast spectroscopy measurements on 1111-type single crystals and thin films~\cite{PhysRevB.81.224504,PhysRevB.87.180509}). In this case, the significantly less coherent itinerant response than in the \pfaofdop\ microcrystal allows for a direct dissociation of the Cooper pairs without the assistance of the mediating boson, as the excess momentum is taken up by the lattice via defects. One may thus expect that the standard Mattis-Bardeen expression for the anomalous skin effect in an impure superconductor with a nodeless gap~\cite{PhysRev.111.412} should apply. Indeed, we find that our experimental data are very well reproduced by this theory (grey line in Fig.~\ref{fig:sc}d). The nodeless character of the superconducting gap is consistent with previous studies of 1111-type compounds~\cite{PhysRevB.87.180509,Andreev_two_gaps_FeSe_2013,Kuzmicheva:2017jo,Charnukha_SFCAO_2015,Charnukha_NdFeAsOF_FSsingularities2015}. The observed agreement between experiment and theory allows us to assign the energy of $33\ \textrm{meV}$ directly to the binding energy of the Cooper pair, $2\Delta$, which results in a gap ratio  $2\Delta/k_\mathrm{B}T_\mathrm{c}\approx7.2$. This value is remarkably similar to that in \pfaofdop\ and, furthermore, to the largest gap ratio identified via ARPES in $\textrm{NdFeAsO}_{1-x}\textrm{F}_x$ (Ref.~\onlinecite{Charnukha_NdFeAsOF_FSsingularities2015}) and optimally doped \bkfaopt\ (Refs.~\onlinecite{PhysRevLett.101.107004,CharnukhaNatCommun2011}). This commonality suggests a single pairing mechanism in all of these compounds and a strong coupling between electrons and the pairing boson. Our work paves the way to future systematic spectroscopic studies of the in-plane infrared charge response of the high-Tc 1111-type iron oxypnictides. Such investigations will enable the extraction of the spectral function of the pairing boson~\cite{PhysRevB.84.174511,Charnukha_NaFeCoAs_OpticsARPES,RevModPhys.62.1027} and its evolution across the phase diagram, shedding light onto the microscopic origin of the highest bulk superconducting transition temperature among the iron-based superconductors.

\begin{acknowledgments}
\par We would like to thank J.~Hilfiker and T.~Tiwald at J.~A.~Woollam~Co for their help with conventional and microfocused ellipsometry measurements and R.~K.~Dumas at Quantum Design for assistance with the magnetic susceptibility measurements on single-crystalline \pfaof. A.C. acknowledges financial support by the Alexander von Humboldt foundation. Work at the University of California, San Diego was supported by Gordon and Betty Moore Foundation.
\end{acknowledgments}




\begin{thebibliography}{50}
\expandafter\ifx\csname natexlab\endcsname\relax\def\natexlab#1{#1}\fi
\expandafter\ifx\csname bibnamefont\endcsname\relax
  \def\bibnamefont#1{#1}\fi
\expandafter\ifx\csname bibfnamefont\endcsname\relax
  \def\bibfnamefont#1{#1}\fi
\expandafter\ifx\csname citenamefont\endcsname\relax
  \def\citenamefont#1{#1}\fi
\expandafter\ifx\csname url\endcsname\relax
  \def\url#1{\texttt{#1}}\fi
\expandafter\ifx\csname urlprefix\endcsname\relax\def\urlprefix{URL }\fi
\providecommand{\bibinfo}[2]{#2}
\providecommand{\eprint}[2][]{\url{#2}}

\bibitem[{\citenamefont{Kamihara et~al.}(2008)\citenamefont{Kamihara, Watanabe,
  Hirano, and Hosono}}]{kamihara}
\bibinfo{author}{\bibfnamefont{Y.}~\bibnamefont{Kamihara}},
  \bibinfo{author}{\bibfnamefont{T.}~\bibnamefont{Watanabe}},
  \bibinfo{author}{\bibfnamefont{M.}~\bibnamefont{Hirano}}, \bibnamefont{and}
  \bibinfo{author}{\bibfnamefont{H.}~\bibnamefont{Hosono}},
  \bibinfo{journal}{J. Am. Chem. Soc.} \textbf{\bibinfo{volume}{130}},
  \bibinfo{pages}{3296} (\bibinfo{year}{2008}).

\bibitem[{\citenamefont{Inosov et~al.}(2011)\citenamefont{Inosov, Park,
  Charnukha, Li, Boris, Keimer, and Hinkov}}]{PhysRevB.83.214520}
\bibinfo{author}{\bibfnamefont{D.~S.} \bibnamefont{Inosov}},
  \bibinfo{author}{\bibfnamefont{J.~T.} \bibnamefont{Park}},
  \bibinfo{author}{\bibfnamefont{A.}~\bibnamefont{Charnukha}},
  \bibinfo{author}{\bibfnamefont{Y.}~\bibnamefont{Li}},
  \bibinfo{author}{\bibfnamefont{A.~V.} \bibnamefont{Boris}},
  \bibinfo{author}{\bibfnamefont{B.}~\bibnamefont{Keimer}}, \bibnamefont{and}
  \bibinfo{author}{\bibfnamefont{V.}~\bibnamefont{Hinkov}},
  \bibinfo{journal}{Phys. Rev. B} \textbf{\bibinfo{volume}{83}},
  \bibinfo{pages}{214520} (\bibinfo{year}{2011}).

\bibitem[{\citenamefont{Zhi-An et~al.}(2008)\citenamefont{Zhi-An, Wei, Jie,
  Wei, Xiao-Li, {Zheng-Cai}, Guang-Can, Xiao-Li, Li-Ling, Fang
  et~al.}}]{Ren_discovery_Sm1111}
\bibinfo{author}{\bibfnamefont{R.}~\bibnamefont{Zhi-An}},
  \bibinfo{author}{\bibfnamefont{L.}~\bibnamefont{Wei}},
  \bibinfo{author}{\bibfnamefont{Y.}~\bibnamefont{Jie}},
  \bibinfo{author}{\bibfnamefont{Y.}~\bibnamefont{Wei}},
  \bibinfo{author}{\bibfnamefont{S.}~\bibnamefont{Xiao-Li}},
  \bibinfo{author}{\bibnamefont{{Zheng-Cai}}},
  \bibinfo{author}{\bibfnamefont{C.}~\bibnamefont{Guang-Can}},
  \bibinfo{author}{\bibfnamefont{D.}~\bibnamefont{Xiao-Li}},
  \bibinfo{author}{\bibfnamefont{S.}~\bibnamefont{Li-Ling}},
  \bibinfo{author}{\bibfnamefont{Z.}~\bibnamefont{Fang}}, \bibnamefont{et~al.},
  \bibinfo{journal}{Chin. Phys. Lett.} \textbf{\bibinfo{volume}{25}},
  \bibinfo{pages}{2215} (\bibinfo{year}{2008}).

\bibitem[{\citenamefont{He et~al.}(2013)\citenamefont{He, He, Zhang, Zhao, Liu,
  Liu, Mou, Ou, Wang, Li et~al.}}]{MonolayerFeSe_65K_2013}
\bibinfo{author}{\bibfnamefont{S.}~\bibnamefont{He}},
  \bibinfo{author}{\bibfnamefont{J.}~\bibnamefont{He}},
  \bibinfo{author}{\bibfnamefont{W.}~\bibnamefont{Zhang}},
  \bibinfo{author}{\bibfnamefont{L.}~\bibnamefont{Zhao}},
  \bibinfo{author}{\bibfnamefont{D.}~\bibnamefont{Liu}},
  \bibinfo{author}{\bibfnamefont{X.}~\bibnamefont{Liu}},
  \bibinfo{author}{\bibfnamefont{D.}~\bibnamefont{Mou}},
  \bibinfo{author}{\bibfnamefont{Y.~B.} \bibnamefont{Ou}},
  \bibinfo{author}{\bibfnamefont{Q.~Y.} \bibnamefont{Wang}},
  \bibinfo{author}{\bibfnamefont{Z.}~\bibnamefont{Li}}, \bibnamefont{et~al.},
  \bibinfo{journal}{Nature Mater.} \textbf{\bibinfo{volume}{12}},
  \bibinfo{pages}{605} (\bibinfo{year}{2013}).

\bibitem[{\citenamefont{Fan et~al.}(2015)\citenamefont{Fan, Zhang, Liu, Yan,
  Ren, Peng, Xu, Xie, Hu, Zhang et~al.}}]{STM_monolayerFeSe-STO_2015}
\bibinfo{author}{\bibfnamefont{Q.}~\bibnamefont{Fan}},
  \bibinfo{author}{\bibfnamefont{W.~H.} \bibnamefont{Zhang}},
  \bibinfo{author}{\bibfnamefont{X.}~\bibnamefont{Liu}},
  \bibinfo{author}{\bibfnamefont{Y.~J.} \bibnamefont{Yan}},
  \bibinfo{author}{\bibfnamefont{M.~Q.} \bibnamefont{Ren}},
  \bibinfo{author}{\bibfnamefont{R.}~\bibnamefont{Peng}},
  \bibinfo{author}{\bibfnamefont{H.~C.} \bibnamefont{Xu}},
  \bibinfo{author}{\bibfnamefont{B.~P.} \bibnamefont{Xie}},
  \bibinfo{author}{\bibfnamefont{J.~P.} \bibnamefont{Hu}},
  \bibinfo{author}{\bibfnamefont{T.}~\bibnamefont{Zhang}},
  \bibnamefont{et~al.}, \bibinfo{journal}{Nature Phys.}
  \textbf{\bibinfo{volume}{11}}, \bibinfo{pages}{946} (\bibinfo{year}{2015}).

\bibitem[{\citenamefont{Shiogai et~al.}(2016)\citenamefont{Shiogai, Ito,
  Mitsuhashi, Nojima, and Tsukazaki}}]{FeSe-STO-MgO_etching_2016}
\bibinfo{author}{\bibfnamefont{J.}~\bibnamefont{Shiogai}},
  \bibinfo{author}{\bibfnamefont{Y.}~\bibnamefont{Ito}},
  \bibinfo{author}{\bibfnamefont{T.}~\bibnamefont{Mitsuhashi}},
  \bibinfo{author}{\bibfnamefont{T.}~\bibnamefont{Nojima}}, \bibnamefont{and}
  \bibinfo{author}{\bibfnamefont{A.}~\bibnamefont{Tsukazaki}},
  \bibinfo{journal}{Nature Phys.} \textbf{\bibinfo{volume}{12}},
  \bibinfo{pages}{42} (\bibinfo{year}{2016}).

\bibitem[{\citenamefont{Wang et~al.}(2016{\natexlab{a}})\citenamefont{Wang, Ma,
  and Xue}}]{0953-2048-29-12-123001}
\bibinfo{author}{\bibfnamefont{L.}~\bibnamefont{Wang}},
  \bibinfo{author}{\bibfnamefont{X.}~\bibnamefont{Ma}}, \bibnamefont{and}
  \bibinfo{author}{\bibfnamefont{Q.-K.} \bibnamefont{Xue}},
  \bibinfo{journal}{Supercond. Sci. Technol.} \textbf{\bibinfo{volume}{29}},
  \bibinfo{pages}{123001} (\bibinfo{year}{2016}{\natexlab{a}}).

\bibitem[{\citenamefont{Johnston}(2010)}]{Johnston_Review_2010}
\bibinfo{author}{\bibfnamefont{D.~C.} \bibnamefont{Johnston}},
  \bibinfo{journal}{Adv. Phys.} \textbf{\bibinfo{volume}{59}},
  \bibinfo{pages}{803} (\bibinfo{year}{2010}).

\bibitem[{\citenamefont{Richard et~al.}(2011)\citenamefont{Richard, Sato,
  Nakayama, Takahashi, and Ding}}]{0034-4885-74-12-124512}
\bibinfo{author}{\bibfnamefont{P.}~\bibnamefont{Richard}},
  \bibinfo{author}{\bibfnamefont{T.}~\bibnamefont{Sato}},
  \bibinfo{author}{\bibfnamefont{K.}~\bibnamefont{Nakayama}},
  \bibinfo{author}{\bibfnamefont{T.}~\bibnamefont{Takahashi}},
  \bibnamefont{and} \bibinfo{author}{\bibfnamefont{H.}~\bibnamefont{Ding}},
  \bibinfo{journal}{Rep. Prog. Phys.} \textbf{\bibinfo{volume}{74}},
  \bibinfo{pages}{124512} (\bibinfo{year}{2011}).

\bibitem[{\citenamefont{Charnukha}(2014)}]{0953-8984-26-25-253203}
\bibinfo{author}{\bibfnamefont{A.}~\bibnamefont{Charnukha}},
  \bibinfo{journal}{J. Phys.: Condens. Matter} \textbf{\bibinfo{volume}{26}},
  \bibinfo{pages}{253203} (\bibinfo{year}{2014}).

\bibitem[{\citenamefont{Dai}(2015)}]{RevModPhys.87.855}
\bibinfo{author}{\bibfnamefont{P.}~\bibnamefont{Dai}}, \bibinfo{journal}{Rev.
  Mod. Phys.} \textbf{\bibinfo{volume}{87}}, \bibinfo{pages}{855}
  (\bibinfo{year}{2015}).

\bibitem[{\citenamefont{Dai et~al.}(2012)\citenamefont{Dai, Hu, and
  Dagotto}}]{NatPhys_Dai_Dagotto_ReviewMagnPnictides}
\bibinfo{author}{\bibfnamefont{P.}~\bibnamefont{Dai}},
  \bibinfo{author}{\bibfnamefont{J.}~\bibnamefont{Hu}}, \bibnamefont{and}
  \bibinfo{author}{\bibfnamefont{E.}~\bibnamefont{Dagotto}},
  \bibinfo{journal}{Nature Phys.} \textbf{\bibinfo{volume}{8}},
  \bibinfo{pages}{709} (\bibinfo{year}{2012}).

\bibitem[{\citenamefont{Lee et~al.}(2014)\citenamefont{Lee, Schmitt, Moore,
  Johnston, Cui, Li, Yi, Liu, Hashimoto, Zhang
  et~al.}}]{Shen_MonolayerFeSe_2014}
\bibinfo{author}{\bibfnamefont{J.~J.} \bibnamefont{Lee}},
  \bibinfo{author}{\bibfnamefont{F.~T.} \bibnamefont{Schmitt}},
  \bibinfo{author}{\bibfnamefont{R.~G.} \bibnamefont{Moore}},
  \bibinfo{author}{\bibfnamefont{S.}~\bibnamefont{Johnston}},
  \bibinfo{author}{\bibfnamefont{Y.~T.} \bibnamefont{Cui}},
  \bibinfo{author}{\bibfnamefont{W.}~\bibnamefont{Li}},
  \bibinfo{author}{\bibfnamefont{M.}~\bibnamefont{Yi}},
  \bibinfo{author}{\bibfnamefont{Z.~K.} \bibnamefont{Liu}},
  \bibinfo{author}{\bibfnamefont{M.}~\bibnamefont{Hashimoto}},
  \bibinfo{author}{\bibfnamefont{Y.}~\bibnamefont{Zhang}},
  \bibnamefont{et~al.}, \bibinfo{journal}{Nature}
  \textbf{\bibinfo{volume}{515}}, \bibinfo{pages}{245} (\bibinfo{year}{2014}).

\bibitem[{\citenamefont{Rebec et~al.}(2017)\citenamefont{Rebec, Jia, Zhang,
  Hashimoto, Lu, Moore, and Shen}}]{PhysRevLett.118.067002}
\bibinfo{author}{\bibfnamefont{S.~N.} \bibnamefont{Rebec}},
  \bibinfo{author}{\bibfnamefont{T.}~\bibnamefont{Jia}},
  \bibinfo{author}{\bibfnamefont{C.}~\bibnamefont{Zhang}},
  \bibinfo{author}{\bibfnamefont{M.}~\bibnamefont{Hashimoto}},
  \bibinfo{author}{\bibfnamefont{D.~H.} \bibnamefont{Lu}},
  \bibinfo{author}{\bibfnamefont{R.~G.} \bibnamefont{Moore}}, \bibnamefont{and}
  \bibinfo{author}{\bibfnamefont{Z.~X.} \bibnamefont{Shen}},
  \bibinfo{journal}{Phys. Rev. Lett.} \textbf{\bibinfo{volume}{118}},
  \bibinfo{pages}{067002} (\bibinfo{year}{2017}).

\bibitem[{\citenamefont{Rademaker et~al.}(2016)\citenamefont{Rademaker, Wang,
  Berlijn, and Johnston}}]{1367-2630-18-2-022001}
\bibinfo{author}{\bibfnamefont{L.}~\bibnamefont{Rademaker}},
  \bibinfo{author}{\bibfnamefont{Y.}~\bibnamefont{Wang}},
  \bibinfo{author}{\bibfnamefont{T.}~\bibnamefont{Berlijn}}, \bibnamefont{and}
  \bibinfo{author}{\bibfnamefont{S.}~\bibnamefont{Johnston}},
  \bibinfo{journal}{New J. Phys.} \textbf{\bibinfo{volume}{18}},
  \bibinfo{pages}{022001} (\bibinfo{year}{2016}).

\bibitem[{\citenamefont{Wang et~al.}(2016{\natexlab{b}})\citenamefont{Wang,
  Nakatsukasa, Rademaker, Berlijn, and Johnston}}]{0953-2048-29-5-054009}
\bibinfo{author}{\bibfnamefont{Y.}~\bibnamefont{Wang}},
  \bibinfo{author}{\bibfnamefont{K.}~\bibnamefont{Nakatsukasa}},
  \bibinfo{author}{\bibfnamefont{L.}~\bibnamefont{Rademaker}},
  \bibinfo{author}{\bibfnamefont{T.}~\bibnamefont{Berlijn}}, \bibnamefont{and}
  \bibinfo{author}{\bibfnamefont{S.}~\bibnamefont{Johnston}},
  \bibinfo{journal}{Supercond. Sci. Technol.} \textbf{\bibinfo{volume}{29}},
  \bibinfo{pages}{054009} (\bibinfo{year}{2016}{\natexlab{b}}).

\bibitem[{\citenamefont{Miyata et~al.}(2015)\citenamefont{Miyata, Nakayama,
  Sugawara, Sato, and Takahashi}}]{Monolayer_FeSe_dopingdepARPES2015}
\bibinfo{author}{\bibfnamefont{Y.}~\bibnamefont{Miyata}},
  \bibinfo{author}{\bibfnamefont{K.}~\bibnamefont{Nakayama}},
  \bibinfo{author}{\bibfnamefont{K.}~\bibnamefont{Sugawara}},
  \bibinfo{author}{\bibfnamefont{T.}~\bibnamefont{Sato}}, \bibnamefont{and}
  \bibinfo{author}{\bibfnamefont{T.}~\bibnamefont{Takahashi}},
  \bibinfo{journal}{Nature Mater.} \textbf{\bibinfo{volume}{14}},
  \bibinfo{pages}{775} (\bibinfo{year}{2015}).

\bibitem[{\citenamefont{Zhigadlo et~al.}(2012)\citenamefont{Zhigadlo, Weyeneth,
  Katrych, Moll, Rogacki, Bosma, Puzniak, Karpinski, and
  Batlogg}}]{PhysRevB.86.214509}
\bibinfo{author}{\bibfnamefont{N.~D.} \bibnamefont{Zhigadlo}},
  \bibinfo{author}{\bibfnamefont{S.}~\bibnamefont{Weyeneth}},
  \bibinfo{author}{\bibfnamefont{S.}~\bibnamefont{Katrych}},
  \bibinfo{author}{\bibfnamefont{P.~J.~W.} \bibnamefont{Moll}},
  \bibinfo{author}{\bibfnamefont{K.}~\bibnamefont{Rogacki}},
  \bibinfo{author}{\bibfnamefont{S.}~\bibnamefont{Bosma}},
  \bibinfo{author}{\bibfnamefont{R.}~\bibnamefont{Puzniak}},
  \bibinfo{author}{\bibfnamefont{J.}~\bibnamefont{Karpinski}},
  \bibnamefont{and} \bibinfo{author}{\bibfnamefont{B.}~\bibnamefont{Batlogg}},
  \bibinfo{journal}{Phys. Rev. B} \textbf{\bibinfo{volume}{86}},
  \bibinfo{pages}{214509} (\bibinfo{year}{2012}).

\bibitem[{\citenamefont{Dubroka et~al.}(2008)\citenamefont{Dubroka, Kim,
  R{\"o}ssle, Malik, Drew, Liu, Wu, Chen, and Bernhard}}]{Dubroka:2008id}
\bibinfo{author}{\bibfnamefont{A.}~\bibnamefont{Dubroka}},
  \bibinfo{author}{\bibfnamefont{K.~W.} \bibnamefont{Kim}},
  \bibinfo{author}{\bibfnamefont{M.}~\bibnamefont{R{\"o}ssle}},
  \bibinfo{author}{\bibfnamefont{V.~K.} \bibnamefont{Malik}},
  \bibinfo{author}{\bibfnamefont{A.~J.} \bibnamefont{Drew}},
  \bibinfo{author}{\bibfnamefont{R.~H.} \bibnamefont{Liu}},
  \bibinfo{author}{\bibfnamefont{G.}~\bibnamefont{Wu}},
  \bibinfo{author}{\bibfnamefont{X.~H.} \bibnamefont{Chen}}, \bibnamefont{and}
  \bibinfo{author}{\bibfnamefont{C.}~\bibnamefont{Bernhard}},
  \bibinfo{journal}{Phys. Rev. Lett.} \textbf{\bibinfo{volume}{101}},
  \bibinfo{pages}{097011} (\bibinfo{year}{2008}).

\bibitem[{\citenamefont{Dong et~al.}(2008)\citenamefont{Dong, Zhang, Xu, Li,
  Li, Hu, Wu, Chen, Dai, Luo et~al.}}]{0295-5075-83-2-27006}
\bibinfo{author}{\bibfnamefont{J.}~\bibnamefont{Dong}},
  \bibinfo{author}{\bibfnamefont{H.~J.} \bibnamefont{Zhang}},
  \bibinfo{author}{\bibfnamefont{G.}~\bibnamefont{Xu}},
  \bibinfo{author}{\bibfnamefont{Z.}~\bibnamefont{Li}},
  \bibinfo{author}{\bibfnamefont{G.}~\bibnamefont{Li}},
  \bibinfo{author}{\bibfnamefont{W.~Z.} \bibnamefont{Hu}},
  \bibinfo{author}{\bibfnamefont{D.}~\bibnamefont{Wu}},
  \bibinfo{author}{\bibfnamefont{G.~F.} \bibnamefont{Chen}},
  \bibinfo{author}{\bibfnamefont{X.}~\bibnamefont{Dai}},
  \bibinfo{author}{\bibfnamefont{J.~L.} \bibnamefont{Luo}},
  \bibnamefont{et~al.}, \bibinfo{journal}{Europhys. Lett.}
  \textbf{\bibinfo{volume}{83}}, \bibinfo{pages}{27006} (\bibinfo{year}{2008}).

\bibitem[{\citenamefont{Boris et~al.}(2009)\citenamefont{Boris, Kovaleva, Seo,
  Kim, Popovich, Matiks, Kremer, and Keimer}}]{boris:027001}
\bibinfo{author}{\bibfnamefont{A.~V.} \bibnamefont{Boris}},
  \bibinfo{author}{\bibfnamefont{N.~N.} \bibnamefont{Kovaleva}},
  \bibinfo{author}{\bibfnamefont{S.~S.~A.} \bibnamefont{Seo}},
  \bibinfo{author}{\bibfnamefont{J.~S.} \bibnamefont{Kim}},
  \bibinfo{author}{\bibfnamefont{P.}~\bibnamefont{Popovich}},
  \bibinfo{author}{\bibfnamefont{Y.}~\bibnamefont{Matiks}},
  \bibinfo{author}{\bibfnamefont{R.~K.} \bibnamefont{Kremer}},
  \bibnamefont{and} \bibinfo{author}{\bibfnamefont{B.}~\bibnamefont{Keimer}},
  \bibinfo{journal}{Phys. Rev. Lett.} \textbf{\bibinfo{volume}{102}},
  \bibinfo{pages}{027001} (\bibinfo{year}{2009}).

\bibitem[{\citenamefont{Mertelj et~al.}(2010)\citenamefont{Mertelj, Kusar,
  Kabanov, Stojchevska, Zhigadlo, Katrych, Bukowski, Karpinski, Weyeneth, and
  Mihailovic}}]{PhysRevB.81.224504}
\bibinfo{author}{\bibfnamefont{T.}~\bibnamefont{Mertelj}},
  \bibinfo{author}{\bibfnamefont{P.}~\bibnamefont{Kusar}},
  \bibinfo{author}{\bibfnamefont{V.~V.} \bibnamefont{Kabanov}},
  \bibinfo{author}{\bibfnamefont{L.}~\bibnamefont{Stojchevska}},
  \bibinfo{author}{\bibfnamefont{N.~D.} \bibnamefont{Zhigadlo}},
  \bibinfo{author}{\bibfnamefont{S.}~\bibnamefont{Katrych}},
  \bibinfo{author}{\bibfnamefont{Z.}~\bibnamefont{Bukowski}},
  \bibinfo{author}{\bibfnamefont{J.}~\bibnamefont{Karpinski}},
  \bibinfo{author}{\bibfnamefont{S.}~\bibnamefont{Weyeneth}}, \bibnamefont{and}
  \bibinfo{author}{\bibfnamefont{D.}~\bibnamefont{Mihailovic}},
  \bibinfo{journal}{Phys. Rev. B} \textbf{\bibinfo{volume}{81}},
  \bibinfo{pages}{224504} (\bibinfo{year}{2010}).

\bibitem[{\citenamefont{Xi et~al.}(2013)\citenamefont{Xi, Dai, Homes, Kidszun,
  Haindl, and Carr}}]{PhysRevB.87.180509}
\bibinfo{author}{\bibfnamefont{X.}~\bibnamefont{Xi}},
  \bibinfo{author}{\bibfnamefont{Y.~M.} \bibnamefont{Dai}},
  \bibinfo{author}{\bibfnamefont{C.~C.} \bibnamefont{Homes}},
  \bibinfo{author}{\bibfnamefont{M.}~\bibnamefont{Kidszun}},
  \bibinfo{author}{\bibfnamefont{S.}~\bibnamefont{Haindl}}, \bibnamefont{and}
  \bibinfo{author}{\bibfnamefont{G.~L.} \bibnamefont{Carr}},
  \bibinfo{journal}{Phys. Rev. B} \textbf{\bibinfo{volume}{87}},
  \bibinfo{pages}{180509} (\bibinfo{year}{2013}).

\bibitem[{\citenamefont{Liu et~al.}(2010)\citenamefont{Liu, Lee, Palczewski,
  Yan, Kondo, Harmon, McCallum, Lograsso, and Kaminski}}]{PhysRevB.82.075135}
\bibinfo{author}{\bibfnamefont{C.}~\bibnamefont{Liu}},
  \bibinfo{author}{\bibfnamefont{Y.}~\bibnamefont{Lee}},
  \bibinfo{author}{\bibfnamefont{A.~D.} \bibnamefont{Palczewski}},
  \bibinfo{author}{\bibfnamefont{J.~Q.} \bibnamefont{Yan}},
  \bibinfo{author}{\bibfnamefont{T.}~\bibnamefont{Kondo}},
  \bibinfo{author}{\bibfnamefont{B.~N.} \bibnamefont{Harmon}},
  \bibinfo{author}{\bibfnamefont{R.~W.} \bibnamefont{McCallum}},
  \bibinfo{author}{\bibfnamefont{T.~A.} \bibnamefont{Lograsso}},
  \bibnamefont{and} \bibinfo{author}{\bibfnamefont{A.}~\bibnamefont{Kaminski}},
  \bibinfo{journal}{Phys. Rev. B} \textbf{\bibinfo{volume}{82}},
  \bibinfo{pages}{075135} (\bibinfo{year}{2010}).

\bibitem[{\citenamefont{Yang et~al.}(2011)\citenamefont{Yang, Xie, Zhou, Zhang,
  Ge, Wu, Wang, Chen, and Feng}}]{Yang2011460}
\bibinfo{author}{\bibfnamefont{L.~X.} \bibnamefont{Yang}},
  \bibinfo{author}{\bibfnamefont{B.~P.} \bibnamefont{Xie}},
  \bibinfo{author}{\bibfnamefont{B.}~\bibnamefont{Zhou}},
  \bibinfo{author}{\bibfnamefont{Y.}~\bibnamefont{Zhang}},
  \bibinfo{author}{\bibfnamefont{Q.~Q.} \bibnamefont{Ge}},
  \bibinfo{author}{\bibfnamefont{F.}~\bibnamefont{Wu}},
  \bibinfo{author}{\bibfnamefont{X.~F.} \bibnamefont{Wang}},
  \bibinfo{author}{\bibfnamefont{X.~H.} \bibnamefont{Chen}}, \bibnamefont{and}
  \bibinfo{author}{\bibfnamefont{D.~L.} \bibnamefont{Feng}},
  \bibinfo{journal}{J. Am. Chem. Soc.} \textbf{\bibinfo{volume}{72}},
  \bibinfo{pages}{460} (\bibinfo{year}{2011}).

\bibitem[{\citenamefont{Charnukha
  et~al.}(2015{\natexlab{a}})\citenamefont{Charnukha, Evtushinsky, Matt, Xu,
  Shi, B{\"u}chner, Zhigadlo, Batlogg, and
  Borisenko}}]{Charnukha_NdFeAsOF_FSsingularities2015}
\bibinfo{author}{\bibfnamefont{A.}~\bibnamefont{Charnukha}},
  \bibinfo{author}{\bibfnamefont{D.~V.} \bibnamefont{Evtushinsky}},
  \bibinfo{author}{\bibfnamefont{C.~E.} \bibnamefont{Matt}},
  \bibinfo{author}{\bibfnamefont{N.}~\bibnamefont{Xu}},
  \bibinfo{author}{\bibfnamefont{M.}~\bibnamefont{Shi}},
  \bibinfo{author}{\bibfnamefont{B.}~\bibnamefont{B{\"u}chner}},
  \bibinfo{author}{\bibfnamefont{N.~D.} \bibnamefont{Zhigadlo}},
  \bibinfo{author}{\bibfnamefont{B.}~\bibnamefont{Batlogg}}, \bibnamefont{and}
  \bibinfo{author}{\bibfnamefont{S.~V.} \bibnamefont{Borisenko}},
  \bibinfo{journal}{Sci. Rep.} \textbf{\bibinfo{volume}{5}},
  \bibinfo{pages}{18273} (\bibinfo{year}{2015}{\natexlab{a}}).

\bibitem[{\citenamefont{Charnukha
  et~al.}(2015{\natexlab{b}})\citenamefont{Charnukha, Thirupathaiah,
  Zabolotnyy, B{\"u}chner, Zhigadlo, Batlogg, Yaresko, and
  Borisenko}}]{Charnukha_SFCAO_2015}
\bibinfo{author}{\bibfnamefont{A.}~\bibnamefont{Charnukha}},
  \bibinfo{author}{\bibfnamefont{S.}~\bibnamefont{Thirupathaiah}},
  \bibinfo{author}{\bibfnamefont{V.~B.} \bibnamefont{Zabolotnyy}},
  \bibinfo{author}{\bibfnamefont{B.}~\bibnamefont{B{\"u}chner}},
  \bibinfo{author}{\bibfnamefont{N.~D.} \bibnamefont{Zhigadlo}},
  \bibinfo{author}{\bibfnamefont{B.}~\bibnamefont{Batlogg}},
  \bibinfo{author}{\bibfnamefont{A.~N.} \bibnamefont{Yaresko}},
  \bibnamefont{and} \bibinfo{author}{\bibfnamefont{S.~V.}
  \bibnamefont{Borisenko}}, \bibinfo{journal}{Sci. Rep.}
  \textbf{\bibinfo{volume}{5}}, \bibinfo{pages}{10392}
  (\bibinfo{year}{2015}{\natexlab{b}}).

\bibitem[{\citenamefont{Takeda et~al.}(2012)\citenamefont{Takeda, Ueda, Takano,
  Yamamoto, and Naito}}]{0953-2048-25-3-035007}
\bibinfo{author}{\bibfnamefont{S.}~\bibnamefont{Takeda}},
  \bibinfo{author}{\bibfnamefont{S.}~\bibnamefont{Ueda}},
  \bibinfo{author}{\bibfnamefont{S.}~\bibnamefont{Takano}},
  \bibinfo{author}{\bibfnamefont{A.}~\bibnamefont{Yamamoto}}, \bibnamefont{and}
  \bibinfo{author}{\bibfnamefont{M.}~\bibnamefont{Naito}},
  \bibinfo{journal}{Supercond. Sci. Technol.} \textbf{\bibinfo{volume}{25}},
  \bibinfo{pages}{035007} (\bibinfo{year}{2012}).

\bibitem[{\citenamefont{Charnukha
  et~al.}(2011{\natexlab{a}})\citenamefont{Charnukha, Dolgov, Golubov, Matiks,
  Sun, Lin, Keimer, and Boris}}]{PhysRevB.84.174511}
\bibinfo{author}{\bibfnamefont{A.}~\bibnamefont{Charnukha}},
  \bibinfo{author}{\bibfnamefont{O.~V.} \bibnamefont{Dolgov}},
  \bibinfo{author}{\bibfnamefont{A.~A.} \bibnamefont{Golubov}},
  \bibinfo{author}{\bibfnamefont{Y.}~\bibnamefont{Matiks}},
  \bibinfo{author}{\bibfnamefont{D.~L.} \bibnamefont{Sun}},
  \bibinfo{author}{\bibfnamefont{C.~T.} \bibnamefont{Lin}},
  \bibinfo{author}{\bibfnamefont{B.}~\bibnamefont{Keimer}}, \bibnamefont{and}
  \bibinfo{author}{\bibfnamefont{A.~V.} \bibnamefont{Boris}},
  \bibinfo{journal}{Phys. Rev. B} \textbf{\bibinfo{volume}{84}},
  \bibinfo{pages}{174511} (\bibinfo{year}{2011}{\natexlab{a}}).

\bibitem[{\citenamefont{Tinkham}(1995{\natexlab{a}})}]{Tinkham_superconductivity_1995}
\bibinfo{author}{\bibfnamefont{M.}~\bibnamefont{Tinkham}},
  \emph{\bibinfo{title}{{Introduction To Superconductivity}}}
  (\bibinfo{publisher}{McGraw-Hill}, \bibinfo{year}{1995}{\natexlab{a}}),
  \bibinfo{edition}{2nd} ed.

\bibitem[{\citenamefont{Basov and Timusk}(2005)}]{RevModPhys.77.721}
\bibinfo{author}{\bibfnamefont{D.~N.} \bibnamefont{Basov}} \bibnamefont{and}
  \bibinfo{author}{\bibfnamefont{T.}~\bibnamefont{Timusk}},
  \bibinfo{journal}{Rev. Mod. Phys.} \textbf{\bibinfo{volume}{77}},
  \bibinfo{pages}{721} (\bibinfo{year}{2005}).

\bibitem[{\citenamefont{Mattis and Bardeen}(1958)}]{PhysRev.111.412}
\bibinfo{author}{\bibfnamefont{D.~C.} \bibnamefont{Mattis}} \bibnamefont{and}
  \bibinfo{author}{\bibfnamefont{J.}~\bibnamefont{Bardeen}},
  \bibinfo{journal}{Phys. Rev.} \textbf{\bibinfo{volume}{111}},
  \bibinfo{pages}{412} (\bibinfo{year}{1958}).

\bibitem[{\citenamefont{Drew et~al.}(2009)\citenamefont{Drew, Niedermayer,
  Baker, Pratt, Blundell, Lancaster, Liu, Wu, Chen, Watanabe
  et~al.}}]{Drew_Bernhard_coexistence_Sm1111_2009}
\bibinfo{author}{\bibfnamefont{A.~J.} \bibnamefont{Drew}},
  \bibinfo{author}{\bibfnamefont{C.}~\bibnamefont{Niedermayer}},
  \bibinfo{author}{\bibfnamefont{P.~J.} \bibnamefont{Baker}},
  \bibinfo{author}{\bibfnamefont{F.~L.} \bibnamefont{Pratt}},
  \bibinfo{author}{\bibfnamefont{S.~J.} \bibnamefont{Blundell}},
  \bibinfo{author}{\bibfnamefont{T.}~\bibnamefont{Lancaster}},
  \bibinfo{author}{\bibfnamefont{R.~H.} \bibnamefont{Liu}},
  \bibinfo{author}{\bibfnamefont{G.}~\bibnamefont{Wu}},
  \bibinfo{author}{\bibfnamefont{X.~H.} \bibnamefont{Chen}},
  \bibinfo{author}{\bibfnamefont{I.}~\bibnamefont{Watanabe}},
  \bibnamefont{et~al.}, \bibinfo{journal}{Nature Mater.}
  \textbf{\bibinfo{volume}{8}}, \bibinfo{pages}{310} (\bibinfo{year}{2009}).

\bibitem[{\citenamefont{Drew et~al.}(2008)\citenamefont{Drew, Pratt, Lancaster,
  Blundell, Baker, Liu, Wu, Chen, Watanabe, Malik et~al.}}]{Drew:2008hp}
\bibinfo{author}{\bibfnamefont{A.~J.} \bibnamefont{Drew}},
  \bibinfo{author}{\bibfnamefont{F.~L.} \bibnamefont{Pratt}},
  \bibinfo{author}{\bibfnamefont{T.}~\bibnamefont{Lancaster}},
  \bibinfo{author}{\bibfnamefont{S.~J.} \bibnamefont{Blundell}},
  \bibinfo{author}{\bibfnamefont{P.~J.} \bibnamefont{Baker}},
  \bibinfo{author}{\bibfnamefont{R.~H.} \bibnamefont{Liu}},
  \bibinfo{author}{\bibfnamefont{G.}~\bibnamefont{Wu}},
  \bibinfo{author}{\bibfnamefont{X.~H.} \bibnamefont{Chen}},
  \bibinfo{author}{\bibfnamefont{I.}~\bibnamefont{Watanabe}},
  \bibinfo{author}{\bibfnamefont{V.~K.} \bibnamefont{Malik}},
  \bibnamefont{et~al.}, \bibinfo{journal}{Phys. Rev. Lett.}
  \textbf{\bibinfo{volume}{101}}, \bibinfo{pages}{097010}
  (\bibinfo{year}{2008}).

\bibitem[{\citenamefont{Martinelli et~al.}(2017)\citenamefont{Martinelli,
  Manfrinetti, Provino, Genovese, Caglieris, Lamura, Ritter, and
  Putti}}]{Martinelli:2017gf}
\bibinfo{author}{\bibfnamefont{A.}~\bibnamefont{Martinelli}},
  \bibinfo{author}{\bibfnamefont{P.}~\bibnamefont{Manfrinetti}},
  \bibinfo{author}{\bibfnamefont{A.}~\bibnamefont{Provino}},
  \bibinfo{author}{\bibfnamefont{A.}~\bibnamefont{Genovese}},
  \bibinfo{author}{\bibfnamefont{F.}~\bibnamefont{Caglieris}},
  \bibinfo{author}{\bibfnamefont{G.}~\bibnamefont{Lamura}},
  \bibinfo{author}{\bibfnamefont{C.}~\bibnamefont{Ritter}}, \bibnamefont{and}
  \bibinfo{author}{\bibfnamefont{M.}~\bibnamefont{Putti}},
  \bibinfo{journal}{Phys. Rev. Lett.} \textbf{\bibinfo{volume}{118}},
  \bibinfo{pages}{055701} (\bibinfo{year}{2017}).

\bibitem[{\citenamefont{Delacr{\'e}taz
  et~al.}(2016)\citenamefont{Delacr{\'e}taz, Gout{\'e}raux, Hartnoll, and
  Karlsson}}]{Delacretaz:2016tg}
\bibinfo{author}{\bibfnamefont{L.~V.} \bibnamefont{Delacr{\'e}taz}},
  \bibinfo{author}{\bibfnamefont{B.}~\bibnamefont{Gout{\'e}raux}},
  \bibinfo{author}{\bibfnamefont{S.~A.} \bibnamefont{Hartnoll}},
  \bibnamefont{and} \bibinfo{author}{\bibfnamefont{A.}~\bibnamefont{Karlsson}}
  (\bibinfo{year}{2016}), \eprint{1612.04381}.

\bibitem[{\citenamefont{Bruin et~al.}(2013)\citenamefont{Bruin, Sakai, Perry,
  and Mackenzie}}]{Bruin:2013hc}
\bibinfo{author}{\bibfnamefont{J.~A.~N.} \bibnamefont{Bruin}},
  \bibinfo{author}{\bibfnamefont{H.}~\bibnamefont{Sakai}},
  \bibinfo{author}{\bibfnamefont{R.~S.} \bibnamefont{Perry}}, \bibnamefont{and}
  \bibinfo{author}{\bibfnamefont{A.~P.} \bibnamefont{Mackenzie}},
  \bibinfo{journal}{Science} \textbf{\bibinfo{volume}{339}},
  \bibinfo{pages}{804} (\bibinfo{year}{2013}).

\bibitem[{\citenamefont{Hayes et~al.}(2016)\citenamefont{Hayes, McDonald,
  Breznay, Helm, Moll, Wartenbe, Shekhter, and Analytis}}]{Hayes:2016bs}
\bibinfo{author}{\bibfnamefont{I.~M.} \bibnamefont{Hayes}},
  \bibinfo{author}{\bibfnamefont{R.~D.} \bibnamefont{McDonald}},
  \bibinfo{author}{\bibfnamefont{N.~P.} \bibnamefont{Breznay}},
  \bibinfo{author}{\bibfnamefont{T.}~\bibnamefont{Helm}},
  \bibinfo{author}{\bibfnamefont{P.~J.~W.} \bibnamefont{Moll}},
  \bibinfo{author}{\bibfnamefont{M.}~\bibnamefont{Wartenbe}},
  \bibinfo{author}{\bibfnamefont{A.}~\bibnamefont{Shekhter}}, \bibnamefont{and}
  \bibinfo{author}{\bibfnamefont{J.~G.} \bibnamefont{Analytis}},
  \bibinfo{journal}{Nature Phys.} \textbf{\bibinfo{volume}{12}},
  \bibinfo{pages}{916} (\bibinfo{year}{2016}).

\bibitem[{\citenamefont{Ferrell and Glover}(1958)}]{PhysRev.109.1398}
\bibinfo{author}{\bibfnamefont{R.~A.} \bibnamefont{Ferrell}} \bibnamefont{and}
  \bibinfo{author}{\bibfnamefont{R.~E.} \bibnamefont{Glover}},
  \bibinfo{journal}{Phys. Rev.} \textbf{\bibinfo{volume}{109}},
  \bibinfo{pages}{1398} (\bibinfo{year}{1958}).

\bibitem[{\citenamefont{Tinkham}(1995{\natexlab{b}})}]{Tinkham_superconductivity_1995_articlestyle}
\bibinfo{author}{\bibfnamefont{M.}~\bibnamefont{Tinkham}},
  \bibinfo{journal}{Introduction To Superconductivity}
  (\bibinfo{year}{1995}{\natexlab{b}}).

\bibitem[{\citenamefont{Nicol et~al.}(1991)\citenamefont{Nicol, Carbotte, and
  Timusk}}]{PhysRevB.43.473}
\bibinfo{author}{\bibfnamefont{E.~J.} \bibnamefont{Nicol}},
  \bibinfo{author}{\bibfnamefont{J.~P.} \bibnamefont{Carbotte}},
  \bibnamefont{and} \bibinfo{author}{\bibfnamefont{T.}~\bibnamefont{Timusk}},
  \bibinfo{journal}{Phys. Rev. B} \textbf{\bibinfo{volume}{43}},
  \bibinfo{pages}{473} (\bibinfo{year}{1991}).

\bibitem[{\citenamefont{Zhang et~al.}(2013)\citenamefont{Zhang, Li, Song, Su,
  Tan, Netherton, Redding, Carr, Sobolev, Schneidewind
  et~al.}}]{PhysRevB.88.064504}
\bibinfo{author}{\bibfnamefont{C.}~\bibnamefont{Zhang}},
  \bibinfo{author}{\bibfnamefont{H.~F.} \bibnamefont{Li}},
  \bibinfo{author}{\bibfnamefont{Y.}~\bibnamefont{Song}},
  \bibinfo{author}{\bibfnamefont{Y.}~\bibnamefont{Su}},
  \bibinfo{author}{\bibfnamefont{G.}~\bibnamefont{Tan}},
  \bibinfo{author}{\bibfnamefont{T.}~\bibnamefont{Netherton}},
  \bibinfo{author}{\bibfnamefont{C.}~\bibnamefont{Redding}},
  \bibinfo{author}{\bibfnamefont{S.~V.} \bibnamefont{Carr}},
  \bibinfo{author}{\bibfnamefont{O.}~\bibnamefont{Sobolev}},
  \bibinfo{author}{\bibfnamefont{A.}~\bibnamefont{Schneidewind}},
  \bibnamefont{et~al.}, \bibinfo{journal}{Phys. Rev. B}
  \textbf{\bibinfo{volume}{88}}, \bibinfo{pages}{064504}
  (\bibinfo{year}{2013}).

\bibitem[{\citenamefont{Charnukha et~al.}(2016)\citenamefont{Charnukha, Post,
  Thirupathaiah, Pr{\"o}pper, Wurmehl, Roslova, Morozov, B{\"u}chner, Yaresko,
  Boris et~al.}}]{Charnukha_NaFeCoAs_OpticsARPES}
\bibinfo{author}{\bibfnamefont{A.}~\bibnamefont{Charnukha}},
  \bibinfo{author}{\bibfnamefont{K.~W.} \bibnamefont{Post}},
  \bibinfo{author}{\bibfnamefont{S.}~\bibnamefont{Thirupathaiah}},
  \bibinfo{author}{\bibfnamefont{D.}~\bibnamefont{Pr{\"o}pper}},
  \bibinfo{author}{\bibfnamefont{S.}~\bibnamefont{Wurmehl}},
  \bibinfo{author}{\bibfnamefont{M.}~\bibnamefont{Roslova}},
  \bibinfo{author}{\bibfnamefont{I.}~\bibnamefont{Morozov}},
  \bibinfo{author}{\bibfnamefont{B.}~\bibnamefont{B{\"u}chner}},
  \bibinfo{author}{\bibfnamefont{A.~N.} \bibnamefont{Yaresko}},
  \bibinfo{author}{\bibfnamefont{A.~V.} \bibnamefont{Boris}},
  \bibnamefont{et~al.}, \bibinfo{journal}{Sci. Rep.}
  \textbf{\bibinfo{volume}{6}}, \bibinfo{pages}{6878} (\bibinfo{year}{2016}).

\bibitem[{\citenamefont{Ponomarev et~al.}(2013)\citenamefont{Ponomarev,
  Kuzmichev, Kuzmicheva, Mikheev, Sudakova, Tchesnokov, Volkova, Vasiliev,
  Pudalov, Sadakov et~al.}}]{Andreev_two_gaps_FeSe_2013}
\bibinfo{author}{\bibfnamefont{Y.~G.} \bibnamefont{Ponomarev}},
  \bibinfo{author}{\bibfnamefont{S.~A.} \bibnamefont{Kuzmichev}},
  \bibinfo{author}{\bibfnamefont{T.~E.} \bibnamefont{Kuzmicheva}},
  \bibinfo{author}{\bibfnamefont{M.~G.} \bibnamefont{Mikheev}},
  \bibinfo{author}{\bibfnamefont{M.~V.} \bibnamefont{Sudakova}},
  \bibinfo{author}{\bibfnamefont{S.~N.} \bibnamefont{Tchesnokov}},
  \bibinfo{author}{\bibfnamefont{O.~S.} \bibnamefont{Volkova}},
  \bibinfo{author}{\bibfnamefont{A.~N.} \bibnamefont{Vasiliev}},
  \bibinfo{author}{\bibfnamefont{V.~M.} \bibnamefont{Pudalov}},
  \bibinfo{author}{\bibfnamefont{A.~V.} \bibnamefont{Sadakov}},
  \bibnamefont{et~al.}, \bibinfo{journal}{J. Supercond. Nov. Magn.}
  \textbf{\bibinfo{volume}{26}}, \bibinfo{pages}{2867} (\bibinfo{year}{2013}).

\bibitem[{\citenamefont{Shamoto et~al.}(2010)\citenamefont{Shamoto, Ishikado,
  Christianson, Lumsden, Wakimoto, Kodama, Iyo, and Arai}}]{PhysRevB.82.172508}
\bibinfo{author}{\bibfnamefont{S.-I.} \bibnamefont{Shamoto}},
  \bibinfo{author}{\bibfnamefont{M.}~\bibnamefont{Ishikado}},
  \bibinfo{author}{\bibfnamefont{A.~D.} \bibnamefont{Christianson}},
  \bibinfo{author}{\bibfnamefont{M.~D.} \bibnamefont{Lumsden}},
  \bibinfo{author}{\bibfnamefont{S.}~\bibnamefont{Wakimoto}},
  \bibinfo{author}{\bibfnamefont{K.}~\bibnamefont{Kodama}},
  \bibinfo{author}{\bibfnamefont{A.}~\bibnamefont{Iyo}}, \bibnamefont{and}
  \bibinfo{author}{\bibfnamefont{M.}~\bibnamefont{Arai}},
  \bibinfo{journal}{Phys. Rev. B} \textbf{\bibinfo{volume}{82}},
  \bibinfo{pages}{172508} (\bibinfo{year}{2010}).

\bibitem[{\citenamefont{Moon et~al.}(2012)\citenamefont{Moon, Schafgans,
  Kasahara, Shibauchi, Terashima, Matsuda, Tanatar, Prozorov, Thaler, Canfield
  et~al.}}]{PhysRevLett.109.027006}
\bibinfo{author}{\bibfnamefont{S.~J.} \bibnamefont{Moon}},
  \bibinfo{author}{\bibfnamefont{A.~A.} \bibnamefont{Schafgans}},
  \bibinfo{author}{\bibfnamefont{S.}~\bibnamefont{Kasahara}},
  \bibinfo{author}{\bibfnamefont{T.}~\bibnamefont{Shibauchi}},
  \bibinfo{author}{\bibfnamefont{T.}~\bibnamefont{Terashima}},
  \bibinfo{author}{\bibfnamefont{Y.}~\bibnamefont{Matsuda}},
  \bibinfo{author}{\bibfnamefont{M.~A.} \bibnamefont{Tanatar}},
  \bibinfo{author}{\bibfnamefont{R.}~\bibnamefont{Prozorov}},
  \bibinfo{author}{\bibfnamefont{A.}~\bibnamefont{Thaler}},
  \bibinfo{author}{\bibfnamefont{P.~C.} \bibnamefont{Canfield}},
  \bibnamefont{et~al.}, \bibinfo{journal}{Phys. Rev. Lett.}
  \textbf{\bibinfo{volume}{109}}, \bibinfo{pages}{027006}
  (\bibinfo{year}{2012}).

\bibitem[{\citenamefont{Kuzmicheva et~al.}(2017)\citenamefont{Kuzmicheva,
  Kuzmichev, Pervakov, Pudalov, and Zhigadlo}}]{Kuzmicheva:2017jo}
\bibinfo{author}{\bibfnamefont{T.~E.} \bibnamefont{Kuzmicheva}},
  \bibinfo{author}{\bibfnamefont{S.~A.} \bibnamefont{Kuzmichev}},
  \bibinfo{author}{\bibfnamefont{K.~S.} \bibnamefont{Pervakov}},
  \bibinfo{author}{\bibfnamefont{V.~M.} \bibnamefont{Pudalov}},
  \bibnamefont{and} \bibinfo{author}{\bibfnamefont{N.~D.}
  \bibnamefont{Zhigadlo}}, \bibinfo{journal}{Phys. Rev. B}
  \textbf{\bibinfo{volume}{95}}, \bibinfo{pages}{094507}
  (\bibinfo{year}{2017}).

\bibitem[{\citenamefont{Li et~al.}(2008)\citenamefont{Li, Hu, Dong, Li, Zheng,
  Chen, Luo, and Wang}}]{PhysRevLett.101.107004}
\bibinfo{author}{\bibfnamefont{G.}~\bibnamefont{Li}},
  \bibinfo{author}{\bibfnamefont{W.~Z.} \bibnamefont{Hu}},
  \bibinfo{author}{\bibfnamefont{J.}~\bibnamefont{Dong}},
  \bibinfo{author}{\bibfnamefont{Z.}~\bibnamefont{Li}},
  \bibinfo{author}{\bibfnamefont{P.}~\bibnamefont{Zheng}},
  \bibinfo{author}{\bibfnamefont{G.~F.} \bibnamefont{Chen}},
  \bibinfo{author}{\bibfnamefont{J.~L.} \bibnamefont{Luo}}, \bibnamefont{and}
  \bibinfo{author}{\bibfnamefont{N.~L.} \bibnamefont{Wang}},
  \bibinfo{journal}{Phys. Rev. Lett.} \textbf{\bibinfo{volume}{101}},
  \bibinfo{pages}{107004} (\bibinfo{year}{2008}).

\bibitem[{\citenamefont{Charnukha
  et~al.}(2011{\natexlab{b}})\citenamefont{Charnukha, Popovich, Matiks, Sun,
  Lin, Yaresko, Keimer, and Boris}}]{CharnukhaNatCommun2011}
\bibinfo{author}{\bibfnamefont{A.}~\bibnamefont{Charnukha}},
  \bibinfo{author}{\bibfnamefont{P.}~\bibnamefont{Popovich}},
  \bibinfo{author}{\bibfnamefont{Y.}~\bibnamefont{Matiks}},
  \bibinfo{author}{\bibfnamefont{D.~L.} \bibnamefont{Sun}},
  \bibinfo{author}{\bibfnamefont{C.~T.} \bibnamefont{Lin}},
  \bibinfo{author}{\bibfnamefont{A.~N.} \bibnamefont{Yaresko}},
  \bibinfo{author}{\bibfnamefont{B.}~\bibnamefont{Keimer}}, \bibnamefont{and}
  \bibinfo{author}{\bibfnamefont{A.~V.} \bibnamefont{Boris}},
  \bibinfo{journal}{Nature Commun.} \textbf{\bibinfo{volume}{2}},
  \bibinfo{pages}{219} (\bibinfo{year}{2011}{\natexlab{b}}).

\bibitem[{\citenamefont{Carbotte}(1990)}]{RevModPhys.62.1027}
\bibinfo{author}{\bibfnamefont{J.~P.} \bibnamefont{Carbotte}},
  \bibinfo{journal}{Rev. Mod. Phys.} \textbf{\bibinfo{volume}{62}},
  \bibinfo{pages}{1027} (\bibinfo{year}{1990}).

\end{thebibliography}
\end{document}